\documentclass[
               amsfonts, amssymb, amsmath,
	       reprint, showkeys,
	       nofootinbib,
	       prd,
	       floatfix, superscriptaddress, natbib,
	      ]{revtex4-2}

\renewcommand\acknowledgements[1]{{\small \emph{Acknowledgements.} #1}}

\usepackage[english]{babel}
\usepackage[utf8]{inputenc}
\usepackage{multirow}
\usepackage{float}
\usepackage[pdftex, pdftitle={Article}, pdfauthor={Author}]{hyperref} 
\usepackage{amsthm}
\usepackage{mathtools}
\usepackage{physics}
\usepackage{xcolor}
\usepackage{graphicx}
\usepackage[left=23mm,right=13mm,top=35mm,columnsep=15pt]{geometry} 
\usepackage{adjustbox}
\usepackage{placeins}
\usepackage[T1]{fontenc}
\usepackage{lipsum}
\usepackage{csquotes}
\usepackage{sidecap}
\usepackage[normalem]{ulem}
\usepackage{enumitem}
\setlist{nosep,leftmargin=\parindent}

\newcommand{\aap}{Astron. Astrophys.}
\newcommand{\apjl}{Astrophys. J. Lett.}
\newcommand{\apjs}{Astrophys. J. Suppl.}
\newcommand{\jcap}{J. Cosmol. Astropart. Phys.}
\newcommand{\mnras}{Mon. Not. Roy. Astron. Soc.}

\newcommand{\pasj}{Publ. Astron. Soc. Jpn.}

\newcommand{\pasp}{Publ. Astron. Soc. Pac.}

\begin{document} 
\title{Cosmological constraints from HSC Y1 lensing convergence PDF}

\author{Leander Thiele}
\email{lthiele@princeton.edu}
\affiliation{Department of Physics, Princeton University, Princeton, NJ 08544, USA}
\affiliation{Center for Data-Driven Discovery, Kavli IPMU (WPI), UTIAS, The University of Tokyo, Kashiwa, Chiba 277-8583, Japan}

\author{Gabriela A. Marques}
\affiliation{Fermi National Accelerator Laboratory, Batavia, IL 60510, USA}
\affiliation{Kavli Institute for Cosmological Physics, University of Chicago, Chicago, IL 60637, USA}
\affiliation{Florida State University, 77 Chieftan Way, Tallahassee, FL 32306, USA}

\author{Jia Liu}
\affiliation{Center for Data-Driven Discovery, Kavli IPMU (WPI), UTIAS, The University of Tokyo, Kashiwa, Chiba 277-8583, Japan}

\author{Masato Shirasaki}
\affiliation{National Astronomical Observatory of Japan (NAOJ), National Institutes of Natural Sciences,
             Osawa, Mitaka, Tokyo 181-8588, Japan}
\affiliation{The Institute of Statistical Mathematics, Tachikawa, Tokyo 190-8562, Japan}

\begin{abstract}
We utilize the probability distribution function (PDF) of normalized convergence maps reconstructed from the Subaru Hyper Suprime-Cam (HSC) Y1 shear catalogue, in combination with the power spectrum, to measure the matter clustering amplitude $S_8 = \sigma_8\sqrt{\Omega_m/0.3}$. The large-scale structure's statistical properties are incompletely described by the traditional two-point statistics, motivating our investigation of the PDF --- a complementary higher-order statistic. By defining the PDF over the standard deviation-normalized convergence map we are able to isolate the non-Gaussian information. We use tailored simulations to compress the data vector and construct a likelihood approximation. We mitigate the impact of survey and astrophysical systematics with cuts on smoothing scales, redshift bins, and data vectors. We find $S_8 = 0.860^{+0.066}_{-0.109}$ from the PDF alone and $S_8 = 0.798^{+0.029}_{-0.042}$ from the combination of PDF and power spectrum ($68\,\%\,\text{CL}$). The PDF improves the power spectrum-only constraint by about $10\,\%$.
\end{abstract}

\maketitle

\section{Introduction}
Weak lensing surveys are rapidly catching up with the
precision afforded by cosmic microwave background (CMB) experiments \cite[e.g.,][]{WMAP9,Planck2015,ACTDR4}.
Weak lensing directly probes the clustering of matter through statistical measurements of distortion of galaxy shapes. 
In recent years, weak lensing surveys have produced increasingly precise
measurements of the matter clustering amplitude $S_8$ \cite[e.g.,][]{Asgari2021,Amon2022,Secco2022,Dalal2023,Li2023},
showing hints of a discrepancy between galaxy lensing and CMB determinations, dubbed the $S_8$-tension. 
Typically, these studies adopt two-point statistics --- the two-point correlation function or the power spectrum.

While two-point statistics are sufficient to describe Gaussian fields such as the CMB,
the large-scale structure that sources the weak lensing signal has undergone nonlinear growth and is thus far from Gaussian.
Therefore, weak lensing non-Gaussian statistics have been proposed to extract complementary information.\footnote{Popular non-Gaussian statistics include minima,
peaks \cite{Jain2000,Marian2009,Dietrich2010,Kratochvil2010,Liu2015,Peel2017,Martinet2018,Harnois-Deraps2021,Sabyr2022},
voids \cite{Davies2021},
Minkowski functionals \cite{Munshi2012,Kratochvil2012,Petri2013,Petri2015,Vicinanza2019,marques2019constraining,Parroni2020},
Betti numbers \cite{Feldbrugge2019,Parroni2021},
persistent homology \cite{Heydenreich2021,Heydenreich2022},
wavelets \cite{Pires2009}, 
scattering transform \cite{Cheng2020,Cheng2021},
moments \cite{Takada2002,Kilbinger2005,VanWaerbeke2013,Petri2013,Petri2015,Porth2021,Gatti2022},
higher-order correlation functions \cite{Cooray2001,Takada2003,Takada2004,Dodelson2005,Huterer2006,Vafaei2010},
density-split statistics \cite{Gruen2018},
and convolutional neural networks \cite{Fluri2019,Fluri2022,Lu2023}.
A comparison of some of these in a forecast setting for Euclid has been carried out by \cite{Euclid2023HOS},
finding about similar performance for many of these statistics but limited to the Fisher approximation.
An even more ambitious attempt are forward-model, field-level methods \citep[e.g.,][]{Porqueres2022,Porqueres2023,Bayer2023,Dai2023,Zurcher2023}}
Equally importantly, non-Gaussian statistics are affected by systematics differently than the two-point statistics.
This is especially relevant for the $S_8$-tension, since whether its origin stems from new physics or systematics is currently under hot debate.
Non-Gaussian statistics will be particularly beneficial for upcoming surveys such as
\emph{Rubin} LSST \cite{Ivezic2019},
Euclid \cite{Laureijs2011},
and \emph{Roman} \cite{Spergel2015}. 
These surveys will provide a high source density and hence probe deeper into the nonlinear regime.

In this work, we focus on the probability distribution function~\cite[PDF,][]{Valageas2000a,Valageas2000b,Munshi2000,Bernardeau2000} of the lensing convergence map\footnote{Also called mass map.}
from Subaru Hyper Suprime-Cam (HSC) Y1 data release~\cite{Aihara2018b,Aihara2018a}.  HSC is currently the deepest large-scale lensing survey and is thus considered a precursor of Stage-IV surveys.
The PDF collects the amplitudes --- but not shapes --- of correlation functions of all orders. Thus, it is a highly non-Gaussian statistics and contains distinct information ``far'' from the power spectrum. Forecasts have shown promise in  tightening the constraints on not only $S_8$ but also the neutrino mass sum and the dark energy equation of
state~\cite{Liu2016,Patton2017,Liu2019,Thiele2020,Boyle2021,Giblin2023}. 
Formally, the PDF and moments are closely related.
Moments can suffer from instabilities due to a few extreme pixels.
This implies that, in practice, with a limited range of convergence values included in the PDF's bins,
the PDF can still be complementary to moments.

Our work is the first to obtain cosmological constraints from the lensing PDF with observations.
While there exist analytic models of the PDF based on large-deviation theory \cite{Reimberg2018,Barthelemy2020a,Barthelemy2020b}
and the halo model \cite{Kainulainen2011a,Kainulainen2011b,Thiele2020}, they are not yet at the level of accuracy and flexibility required to incorporate the complex survey configurations and systematics. 
Therefore, we use a large set of cosmological simulations tailored to HSC Y1 data to model the lensing PDF and its likelihood. 

\section{Methods}
In this section, we provide a brief overview of the HSC Y1 data, the PDF and power spectrum measurements, the simulations suite, the likelihood, our blinding procedure, and systematics.
More detail may be found in our companion paper on counts of lensing peaks and minima~\cite{GAM2023}.

\subsection{HSC Y1 Data}
We use lensing convergence maps estimated from the HSC Y1 galaxy shapes catalogue~\cite{Mandelbaum2018a}.
After applying masks, the shear maps span $136.9\,\text{deg}^2$ in six spatially disjoint fields. We adopt galaxy redshifts determined using the \texttt{MLZ} code~\cite{Tanaka2018}.
After applying redshift cuts of $0.3<z_s<1.5$ and restricting to sources with reliable shape measurements, we obtained a total number density of $\sim 17\,\text{arcmin}^{-2}$.
We split source galaxies into four tomographic redshift bins with edges [0.3, 0.6, 0.9, 1.2, 1.5]
and construct convergence maps via Kaiser-Squires inversion~\cite{Kaiser1993}.

\subsection{PDF and power spectrum}
\label{sec:datavec}
\begin{figure}
\includegraphics[width=\linewidth]{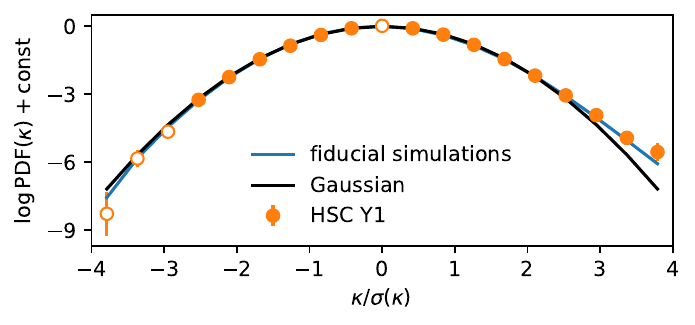}
\caption{Example of the non-Gaussian lensing convergence PDFs from the HSC Y1 data (orange) and the fiducial simulation (blue), for the tomographic redshift bin $0.6<z_s<0.9$ and with a smoothing scale $\theta_s = 5\,\text{arcmin}$. For most of the data points the error bars are invisible. Unfilled data points are not used in the analysis. The PDF's deviation from a Gaussian (black) provides the cosmological information to be exploited in this work.}
\label{fig:pdf}
\end{figure}
The primary summary statistic we consider is the lensing PDF, calculated as histograms of pixels in normalized convergence maps.
One crucial step in our computation is that we define the PDF over the signal-to-noise ratio, i.e., each convergence map is divided by its standard deviation before its pixels are histogrammed.
This is important to remove information duplicated in the power spectrum.
The PDF's non-Gaussian character becomes apparent in Fig.~\ref{fig:pdf}, where the tails deviate from a Gaussian
(which is dominated by galaxy shape noise).

We measure PDFs on maps with $0.88\,\text{arcmin}$-sidelength pixels smoothed with Gaussian filters $\propto\exp(-\theta^2/\theta_s^2)$,
where we choose $\theta_s=\{5, 7, 10\}\,\text{arcmin}$ to mitigate systematics while retaining non-Gaussian information.
A Gaussian filter has optimal joint localization in configuration and reciprocal space
and is thus a common choice~\citep[e.g.,][]{Sabyr2022,Lu2023,Euclid2023HOS}.
The presence of masking (where we set shear to zero) means that the filter may decrease the signal-to-noise ratio
slightly, but no biases can arise since the mask is consistently implemented in data and simulations.
For each smoothing scale, we histogram in 19 equally spaced bins between -4 and 4 (cf. Fig.~\ref{fig:pdf}).
Of these, we remove the first three bins to minimize contamination by baryons according to tests described in
Sec.~\ref{sec:syst}~\cite[also cf.][]{Sunseri2023}.
Finally, we remove the 10th bin as otherwise the linear constraint from fixing the first three moments would render the PDF's auto-covariance almost singular.

We also consider the auto-power spectrum for each tomographic redshift bin,
deconvolving the survey mask using the pseudo-$C_\ell$ method as implemented in \texttt{NaMaster}~\cite{hivon2002master,alonso2019unified}.
We measure $C_\ell^{\kappa\kappa}$ in 14 logarithmically spaced bins in angular multipole $81<\ell<6580$. Of these, we remove $\ell<285$ as Ref.~\cite{Oguri2018} found unmodeled systematic
errors for these scales in HSC Y1 data. Furthermore, we remove $\ell>1000$ due to possible contamination by baryons. We note that $\ell_\text{max} \sim 1000$ is also comparable to the minimum smoothing scale of $5\,\text{arcmin}$ used in the PDF data vector. This leaves us with four $\ell$-bins.
In contrast to previous two-point analyses \cite{Hikage2019,Hamana2020} we do not include cross-spectra between tomographic redshift bins. 
Upon unblinding the power spectrum-only posteriors we found that the highest tomographic redshift bin causes significant shifts in $S_8$.\footnote{This may be due to deficiencies in our simulations; for example, the simulations use lower resolution at high redshifts which may be not sufficient.
Alternatively, the highest source redshift bin may be subject to photo-$z$ calibration errors~\cite{Dalal2023, Li2023}.}
Thus, we exclude the highest redshift bin in our analysis. 

In summary, our raw data vector consists of
the PDF with 135 dimensions (3 smoothing scales $\times$~3 redshifts $\times$~15 bins)
and the power spectrum with 12 dimensions (3 redshifts $\times$~4 bins).

\subsection{Simulations}
We adopt two sets of $N$-body simulations in our analysis:
\begin{enumerate}
    \item To model the covariance, we use 2268 realizations of the HSC Y1 footprint at a fiducial point
    ($S_8=0.791$, $\Omega_m=0.279$),
    generated from the 108 quasi-independent~\cite{Petri2016} full-sky maps of Ref.~\cite{Takahashi2017}.
    \item To model our statistics, we use simulations at 100 different cosmological models with varying values of $S_8$ and $\Omega_m$, as
    introduced in Ref.~\cite{Shirasaki2021}.
   For each cosmology, 50 quasi-independent realizations are generated by randomly placing observers within the periodic box.
   These simulations are used to construct emulators for the mean theory prediction; thus, they are designed to have better accuracy than the fiducial simulations.
\end{enumerate}
For each of these 7768 (=2268+100$\times$50) realizations, mock galaxy shapes are generated following Refs.~\cite{Shirasaki2014,Shirasaki2017,Shirasaki2019}. Convergence map generation and summary statistic measurements are performed on these simulated mocks identically to as done for the real data.

\subsection{Likelihood and inference}
\label{sec:like}
For the PDF, to reduce the size of the data vector and Gaussianize its likelihood,
we score-compress the logarithmic PDF under the approximation of a Gaussian likelihood~\cite[MOPED,][]{Heavens2000}.\footnote{
	We note that upon combining the PDF with the power spectrum, the compression is likely sub-optimal;
	future work could investigate more rigorous ways to preserve a maximum of information.
	Of course, the choice of likelihood to compute the score cannot induce biases, only information loss.}
This reduces the number of PDF bins from 135 to only 2, corresponding to the number of parameters ($S_8$ and $\Omega_m$).
We then construct at each of the 100 cosmological models a 2$\times$2 covariance matrix, using the 50 realizations.
Finally, we build emulators of both the compressed PDF and the cosmology-dependent inverse covariance using Gaussian processes.\footnote{
	Analytic methods to compute the covariance have been presented in Refs.~\cite{Thiele2020,Uhlemann2023}.}
If we use a cosmology-independent covariance the $S_8$-posterior tightens and its peak is almost unchanged.
However, with that choice the ranks plot discussed below would indicate an overconfident posterior.

The power spectrum data vector is small enough and its distribution is known to be close to Gaussian. Therefore, we apply no data compression to the power spectrum. We build a Gaussian process emulator to model the power spectrum and estimate a cosmology-independent covariance matrix using the fiducial simulations. To jointly analyze the compressed PDF and the power spectrum, we approximate the
cross-correlation between them as constant and estimate it from the fiducial simulations.

We adopt uniform priors on our parameters $S_8$ and $\Omega_m$, in intervals $[0.5, 1.0]$ and $[0.2, 0.4]$, respectively.
Our prior is well-covered by the available simulations.
Markov chain Monte Carlo (MCMC) sampling is performed using \texttt{emcee}~\cite{Foreman-Mackey2013a,Foreman-Mackey2013b}.

\begin{figure}
\includegraphics[width=\linewidth]{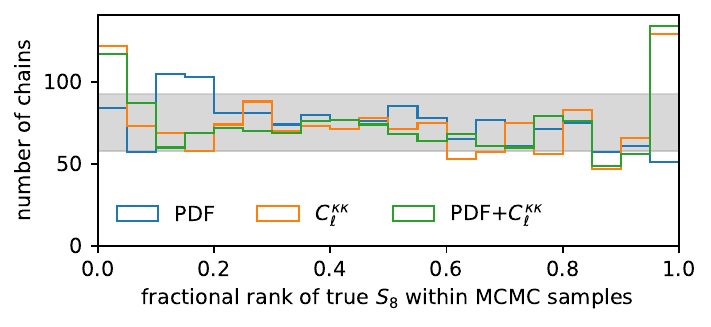}
\caption{Validations of our likelihoods. The histograms are over hundreds of Markov chains, run with realizations from the cosmo-varied simulation set as ``observation''. The $x$-axis is the position of the simulation's true $S_8$ value in the ordered chain. A uniform distribution is expected for a valid likelihood.
The grey band represents the $2\sigma$ interval approximating the bin counts as Poissonian.}
\label{fig:ranks}
\end{figure}

We validate our likelihoods with a ``ranks plot'' \cite[e.g.,][]{Vehtari2019} shown in Fig.~\ref{fig:ranks}.
The rank statistic is a simple calibration test which relies on the fact that for a valid likelihood
random draws from the posterior should be statistically indistinguishable from the true parameter vector.
To construct the ranks plot, we run MCMC on realizations drawn from the 30 cosmology-varied simulations within our prior.
To cleanly separate our training and test sets, all emulators are rebuilt without the test cosmology.
We then order each Markov chain by its $S_8$ values and find the rank of the true $S_8$.
For a valid likelihood, it should be impossible to statistically distinguish the true $S_8$
from randomly selected Monte Carlo samples,
so the ranks should follow a uniform distribution.
If the posterior is over-confident (under-confident), the histogram exhibits a U-shape (inverse U).
However, because we only have a relatively small number of simulated cosmologies available within our prior,
a perfectly uniform distribution may not be possible to attain.
Indeed, Fig.~\ref{fig:ranks} shows an approximately uniform distribution for all data vector choices (PDF, power spectrum, and them jointly),
except for spikes at the edges for the power spectrum and joint statistic.
These spikes are attributable to a small number of simulations near the prior boundary where the power spectrum emulation appears to work less well.
We do not expect these to affect our results as they have little overlap with our final posterior and the problem is mostly due to underestimated
tails which are less important for the confidence levels typically considered.

\subsection{Systematics}
\label{sec:syst}
\begin{figure}
\includegraphics[height=4.7cm]{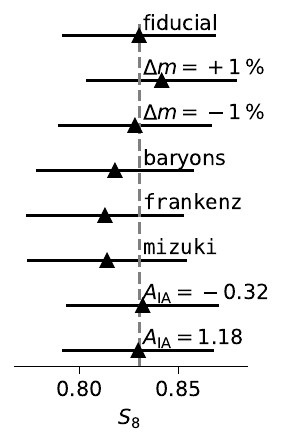}
\includegraphics[height=4.7cm]{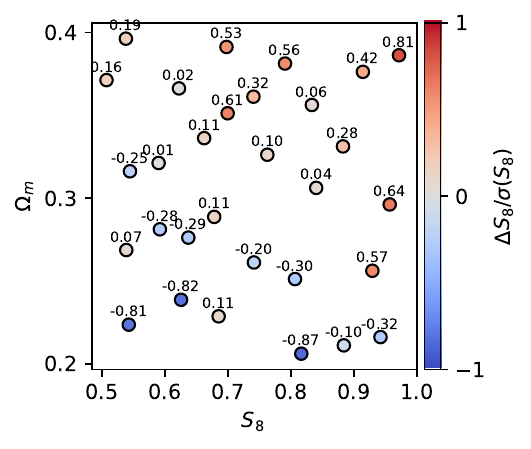}
\caption{\emph{Left}: Estimate of systematic errors.
         We contaminate the PDF+$C_\ell^{\kappa\kappa}$ data vector with various systematics as described
	 in the text. No systematic exceeds $0.45\sigma(S_8)$.
	 \emph{Right}: Assessment of training set and emulator quality.
	 Each point is one of the cosmologies for which cosmo-varied realizations exist.
	 The bias indicated is estimated by averaging over chains run on 25 realizations,
	 while the emulator is trained with the other 25 realizations and without including the cosmology
	 where the inference is performed.
	}
\label{fig:syst}
\end{figure}
We study potential biases caused by effects not included or wrongly described in the simulated model
by running inference on fiducial mocks contaminated with the following methods:
\begin{itemize}
    \item We simulate miscalibration of multiplicative bias by shifting it by $\pm 1\,\%$, corresponding to the uncertainty on the mean quoted in Ref.~\cite{Mandelbaum2018c}.
    \item We assess the impact of photometric redshift uncertainty by generating fiducial mocks with source redshifts from two other codes, \texttt{frankenz} and \texttt{mizuki}.
    \item We model the impact of baryonic effects using the $\kappa$TNG simulations~\cite{Osato2021} --- two sets of convergence maps with and without baryons.
    We contaminate the fiducial mocks with the ratio of hydrodynamic to dark matter-only data vectors.
    \item We estimate the impact of intrinsic alignments similarly to the baryons case by multiplying the fiducial data vectors with a fixed ratio which we obtain from mocks with nonlinear intrinsic alignments~\cite{Harnois-Deraps2022}. We choose alignment amplitudes, $A_\text{IA}=\{ -0.32, 1.18 \}$, comparable to the $1\sigma$ constraints in Refs.~\cite{Hikage2019,Hamana2020}.
\end{itemize}
We show the resulting biases on $S_8$ for these systematics in the left panel of Fig.~\ref{fig:syst} (also see the more comprehensive discussion in Ref.~\cite{GAM2023}).
The most constraining joint PDF+$C_\ell^{\kappa\kappa}$ data vector is used. We do not find biases exceeding $0.45\sigma$ in $S_8$.

In addition, we also consider potential biases caused by imperfections in the simulations and emulators:
\begin{itemize}
    \item The mean recovered $S_8$ on the fiducial simulations is about $1\sigma$ higher than the input, indicating a systematic difference with the cosmology-varied simulations.
    This is likely due to the lower fidelity of the fiducial simulations. With this in mind, the observed $1\sigma$ bias can be considered an upper bound on biases caused by resolution effects in the $N$-body simulations.
    \item To assess biases caused by the limited training set size, at each cosmology we divide the realizations into 25 samples each for training and testing.
    We match the realization indices (i.e., observer positions) between cosmologies so as to maximize statistical independence between training and testing samples.
    In addition, all realizations at the test cosmology are removed from the training. The right panel of Fig.~\ref{fig:syst} shows the average biases on $S_8$ within our prior range, in units of the standard deviation.
    The bias is computed using the mode of the marginalized $S_8$ posterior, to minimize the effect of the prior which can shift the mean.
    The bias is a few ten percent for most of the points, except at the edges where the emulator quality decreases.
    Since the described test aggressively reduces the training set size, our bias estimates are likely a conservative upper bound.
\end{itemize}

\subsection{Blinding}
To help build best practice for non-Gaussian statistical analyses with Stage-IV surveys~\cite[e.g.,][]{Euclid2023HOS}, we follow a three-step blinding procedure in our analysis:
\begin{enumerate}
    \item We build the mock generation pipeline using one random realization from the fiducial model as ``observation''.
    This includes shear bias correction, convergence map reconstruction, masking, in-painting, and summary statistics measurements.
    \item We construct the inference pipeline using the ranks plot and similar tests.
    This includes data compression, covariances, emulators, and MCMC sampling.
    \item We select smoothing scales, redshift bins, and cuts on data vectors to minimize the impact of systematics.
    \item First unblinding: B-modes. We compare the power spectra and PDFs measured in B-mode maps of the HSC Y1 data and our fiducial simulation.\footnote{B-mode maps are built by rotating galaxy shapes by 45 degrees.} We validate that the two distributions are statistically consistent.
    \item Second unblinding: power spectrum. We apply inference to the measured HSC Y1 power spectrum and compare our results internally between different redshift bins and to the official HSC Y1 analyses by Refs.~\cite{Hikage2019,Hamana2020}. We discover systematics in the highest tomographic redshift bin and remove it from our final analysis.
    \item Third unblinding: PDF. We unblind the PDF-only posterior during an invited presentation at the HSC weak lensing working group telecon.
    The initial posterior was approximately uniform. This failure was due to a bug in computing the standard deviations of the HSC data maps.\footnote{
    	Thanks to the feedback from the HSC weak lensing working group, we learned a valuable lesson of comparing the PDFs before moving on to inferences.
	They also pointed out that a more sophisticated blinding policy, as typically employed by collaborations, would have flagged this issue before unblinding.}
    The bug did not affect our previous steps, so once we resolved it we did not change any other parts of the pipeline.
    \item After the initial submission of this paper, we learned that there are two additional multiplicative bias parameters
          arising from galaxy size selection and redshift-dependent responsivity corrections.
          Post-unblinding, we updated our analysis to account for this total multiplicative bias.
          The correction is very small and hence did not change our conclusions.
          However, it made our power spectrum-only results more consistent with the official analysis~\cite{Hikage2019}.
\end{enumerate}
Since the HSC Y1 data is already public
our blinding procedure is rather an honor system.\footnote{During our study, we also became aware of other recently completed non-Gaussian statistical analyses using the same dataset~\cite{Lu2023,LiuX2023}. To avoid confirmation bias, we refrained from reading these papers until we finalized our results.}

\section{Results}
\begin{figure}
\includegraphics[width=\linewidth]{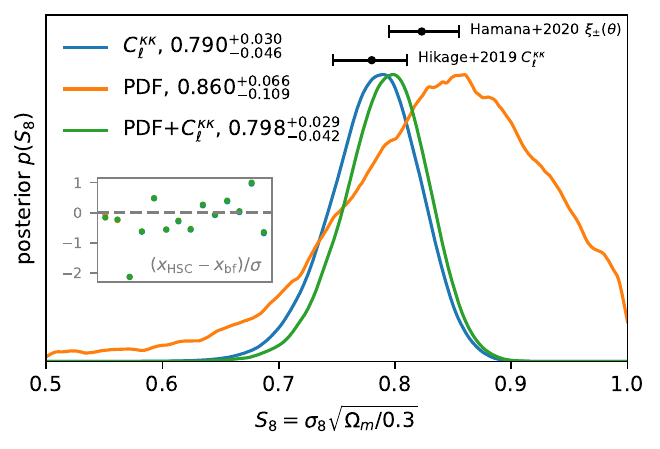}
\caption{$S_8$ posteriors from different data vector choices.
         Since the posteriors are slightly asymmetric, we quote their modes
	 as well as the symmetric $68\,\%$ confidence interval.
         For comparison, we include $68\,\%$ intervals from Refs. \cite{Hikage2019,Hamana2020}.
	 Note that these analyses differ in various details.
	 The inset compares the best-fit model to the data ($\text{PTE}=0.86$).
	 We remind the reader that the PDF does not contain information on the variance.
	}
\label{fig:posterior}
\end{figure}
We show the $S_8$ posteriors obtained from the power spectrum, the PDF, and the two combined in Fig.~\ref{fig:posterior}. In all cases, the constraint on $\Omega_m$ is prior-dominated and thus not shown here for brevity.
For reference, we also show the $68\,\%$ intervals from the official HSC Y1 two-point analyses with the power spectrum~\cite{Hikage2019} and the two-point correlation function~\cite{Hamana2020}.

Our two-point result (blue) agrees well with the previous analyses in terms of mean value. However, our posterior is  wider: $\sigma \simeq 0.042$ compared to $\sigma \simeq 0.03$ in the previous two-point analyses. This is expected as these analyses differ in many respects, such as additional cross-spectra between tomographic redshift bins, choice of priors,
treatment of systematic errors, data vector binning, and scale cuts.

Our PDF-only posterior (orange) shows a clear detection of $S_8$, or specifically \textit{nonlinear} clustering, at $9\,\sigma$. This is after removing variance information from the PDF and thus almost independent of the two-point results. The mean $S_8$ value is slightly higher than the one from $C_\ell^{\kappa\kappa}$ but statistically consistent. Upon combining the PDF and
$C_\ell^{\kappa\kappa}$, we observe a slight upwards shift in the mean $S_8$ and a $\approx\,10\,\%$ tightening of the posterior. This shows that the PDF indeed contains complementary information to the power spectrum. 

In many aspects, our results match the CNN analysis from Ref.~\cite{Lu2023},
including the $S_8$ posterior and an $\mathcal{O}(10\,\%)$ improvement compared to the power spectrum.

\section{Conclusions}
In this work, we obtained the first cosmological constraints using the probability distribution function of weak lensing convergence.
We built convergence maps using the HSC Y1 shear catalogue, constructed a likelihood based on tailored numerical simulations, and validated that known survey and astrophysical systematics are under control after applying cuts on smoothing scales, redshift, and data vectors. We designed and followed a three-step blinding procedure to minimize confirmation bias. We make our inference code publicly available at \href{https://github.com/leanderthiele/HSC_Y1_kappapdf_production}{this URL}.

The PDF improves the power spectrum-only constraint by about $10\,\%$. For the clustering amplitude, we obtained $S_8 = 0.860^{+0.066}_{-0.109}$ and $S_8 = 0.798^{+0.029}_{-0.042}$ from the lensing PDF alone and the combination of PDF and power spectrum, respectively ($68\,\%\,\text{CL}$). We computed the PDF on convergence maps normalized by their standard deviation, hence maximally removed two-point information from the PDF. Our results are consistent with previous analyses on the same data and show that the PDF provides additional information not contained in the power spectrum.

We find no tension between the $S_8$ inferred from HSC Y1 lensing and from primordial CMB measurements.

Future work could investigate alternatives to the probably lossy compression step
as well as the forward modeling of systematics to allow inclusion of smaller scales.
Additionally, Stage-IV data will be sensitive to cosmological parameters beyond $S_8$, for example, the neutrino mass sum.
The PDF could be instrumental in complementing two-point statistics in order to optimally constrain such model extensions.

\acknowledgements{
We thank Sihao Cheng, Will Coulton, Daniela Grand\'on, Kevin Huffenberger, Xiangchong Li, Surhud More, Ken Osato, David Spergel, Sunao Sugiyama, and Masahiro Takada for useful discussions. We thank Joachim Harnois-D\'eraps for sharing the IA mocks. 
We thank the organizers of the Kyoto CMBxLSS workshop, during which this work was completed.
The work of LT is supported by the NSF grant AST~2108078.
This work was supported by JSPS KAKENHI Grants 23K13095 and  23H00107 (to JL), 	
19K14767 and 20H05861 (to MS). This work was supported in part by the UTokyo-Princeton Strategic Partnership Teaching and Research Collaboration.
This manuscript has been authored by Fermi Research Alliance, LLC under Contract No. DE-AC02-07CH11359 with the U.S. Department of Energy, Office of Science, Office of High Energy Physics.
The authors are pleased to acknowledge that the work reported on in this paper was substantially performed using the Princeton Research Computing resources at Princeton University which is consortium of groups led by the Princeton Institute for Computational Science and Engineering (PICSciE) and Office of Information Technology's Research Computing.\
}

\appendix

\section{Information}

\begin{figure}
\includegraphics[width=\linewidth]{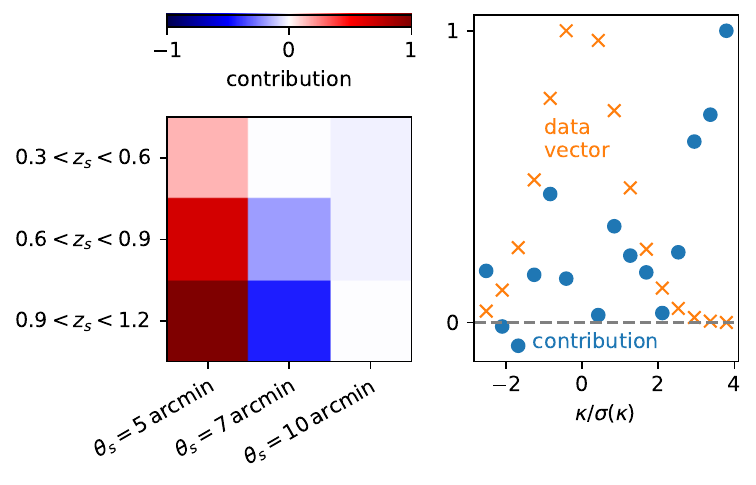}
\caption{Illustration of the compression weights for the PDF.
         We picked a high-$S_8$ cosmology (with close to fiducial $\Omega_m$) from our simulations.
         By taking the element-wise product between the first MOPED compression vector
	 (corresponding to $S_8$) and the data vector
	 we can get intuition as to where information is coming from.
	 For this figure, we subtract the corresponding product for the fiducial simulations.
	 Note that the data vector shown in orange is just for illustration and has been rescaled
	 and shifted for convenience.\\
	 \emph{Left}: contributions from different source redshifts and smoothing scales
	 (with the $\kappa$-bins summed over).
	 \emph{Right}:  contributions from different $\kappa$-bins with source redshifts
	 and smoothing scales summed over.
	}
\label{fig:moped}
\end{figure}

In Fig.~\ref{fig:moped}, we illustrate where the information in the PDF is primarily coming from.

In the left panel, we focus on redshift bins and smoothing scales.
We observe that, anologously to the two-point statistics, the highest redshift bin is the most informative.
Furthermore, the largest smoothing scale ($10\,\text{arcmin}$) is quite uninformative,
due to the nearly Gaussian character of the field.
On the other hand, the smaller smoothing scales contain information and the compression appears
to use a difference between them.

In the right panel, we focus on the convergence bins.
Even though it might be expected that the higher signal-to-noise ratio close to the PDF's peak
would make these bins the most informative,
we observe that the high-$\kappa$ tail is actually where most of the information is coming from.
This is consistent with our picture in Fig.~\ref{fig:pdf} that the deviations from Gaussianity
in the tail are the focus of this work.

\section{Data vector choices}

\begin{figure}
\includegraphics[width=\linewidth]{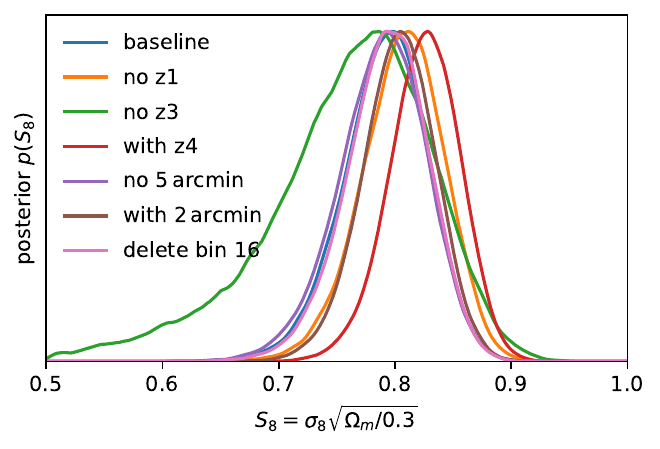}
\caption{The effect of various cuts and extensions in the data vector on the final posterior.
         All posteriors plotted are for $\text{PDF}+C_\ell^{\kappa\kappa}$, the blue baseline
	 being the same curve as in Fig.~\ref{fig:posterior}.\\
        }
\label{fig:cuts}
\end{figure}

In Fig.~\ref{fig:cuts}, we eplore the effects of various modifications to the data vector.
First, we remove the lowest and highest tomographic redshift bin (orange, green) and
observe that the higher one contributes much more signal-to-noise.
Then, we add the tomographic redshift bin that was excluded in our baseline analysis,
c.f. Sec.~\ref{sec:datavec} (red). As discussed in the text, including this bin leads
to a substantial shift in the posterior.\\
The last three tests apply only to the PDF part of the data vector.
First, we remove the smallest smoothing scale (purple).
Then we add a $2\,\text{arcmin}$ smoothing scale (brown).
We observe that using the smaller smoothing scale tightens the posterior substantially,
indicating that there is useful information. However, in our baseline analysis
such small scales are not used due to concerns about baryonic systematics.
Finally, we delete the 17th convergence bin (instead of the 10th, c.f. Sec.~\ref{sec:datavec}),
and find that this modification of the data vector does not alter the posterior.

\section{Likelihood choices}

\begin{figure}
\includegraphics[width=\linewidth]{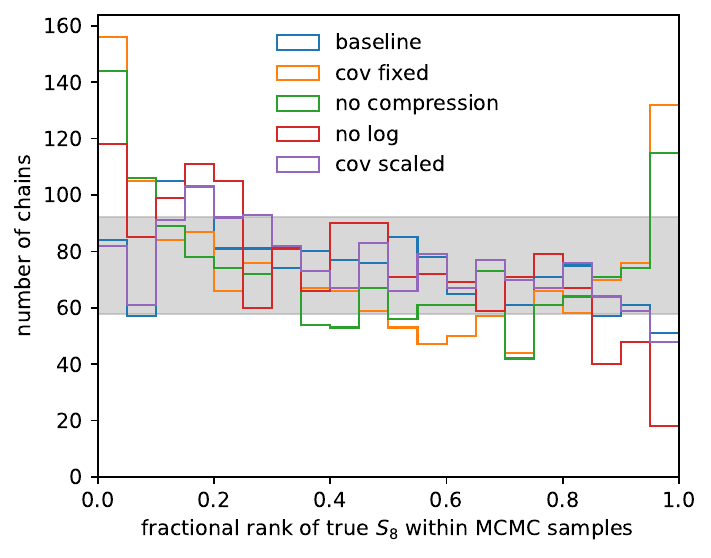}
\includegraphics[width=\linewidth]{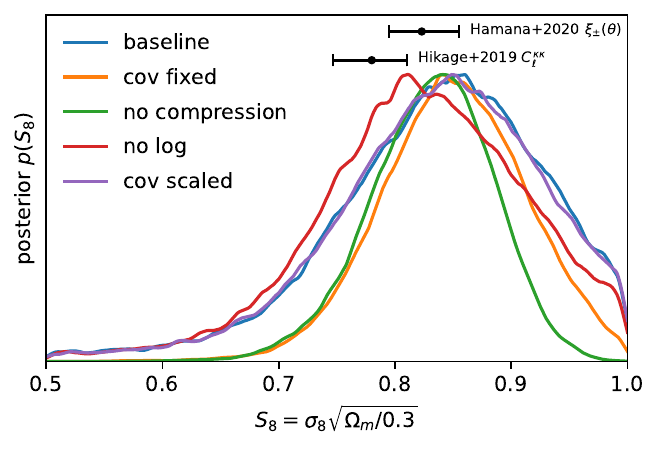}
\caption{The effect of changing the PDF likelihood in various ways (c.f. Sec.~\ref{sec:like}).
         In all plots, the data vector considered is PDF only.
	 The top panel is the ranks plot, c.f. Fig.~\ref{fig:ranks},
	 and the bottom panel are the posteriors on real data obtained with the different likelihoods.
	 We observe that the baseline likelihood used in the main part of the paper
	 exhibits the healthiest calibration diagnostics.
	 Furthermore, we see that the different likelihood assumptions lead to visible but mild
	 shifts in the posterior.
        }
\label{fig:miscalibration}
\end{figure}

In Fig.~\ref{fig:miscalibration}, we show what happens to the PDF-only posterior
when various choices in our likelihood construction are altered.
As shown on the upper panel, the baseline used in the main text (blue)
exhibits the best rank statistics (closest to uniform distribution).
The modifications deviate from uniform distribution.
The resulting posteriors are still broadly consistent.

\section{Tomography}

\begin{figure}
\includegraphics[width=\linewidth]{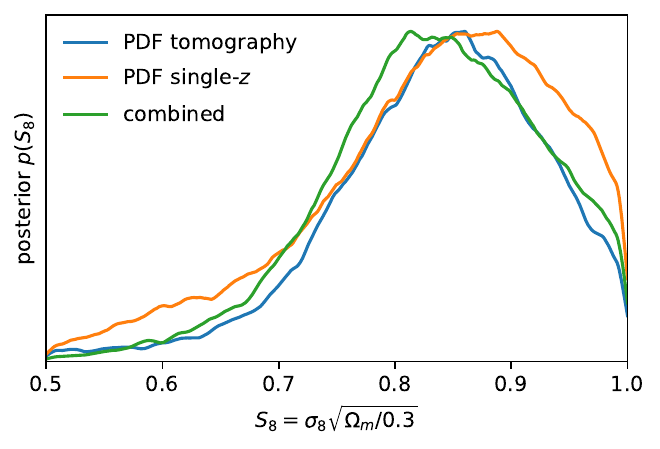}
\caption{
	 Impact of tomography in the PDF-only analysis.
	 We consider only the lower three redshift bins, as explained in Sec.~\ref{sec:datavec}.
	 We observe that the posterior obtained with tomography, blue, is tighter than
	 the one without, orange.
	 The combined posterior, green, is consistent with the individual ones but slightly broader
	 then the tomographic result, presumably due to small inefficiencies in the compression.
	}
\label{fig:tomography}
\end{figure}

For two-point analyses, including the auto- and cross-power spectra from tomographic redshift bins
saturates the information.
The situation is more complicated with non-Gaussian statistics because the shape noise
propagates in a non-linear way into the data vector and posteriors.
In the main text, we used a tomographic analysis,
motivated by previous work that indicated benefits to tomography~\cite{Liu2019}.
In Fig.~\ref{fig:tomography}, we show what happens when we collapse all sources in a single bin (orange).
We observe that the tomographic analysis does yield a somewhat tighter posterior.
Of course, one can also combine tomographic and non-tomographic data vectors,
or use various other mergers of redshift bins.
We do not expect such strategies to significantly affect the constraining power in our case,
but more generally they warrant further study.

\end{document}